\documentclass[prl,amssymb,amsfonts,twocolumn,showpacs]{revtex4}

  \def\beq{\begin{equation}}
  \def\eeq{\end{equation}}
  \def\bea{\begin{eqnarray}}
\def\eea{\end{eqnarray}}
  \def\6{\langle}
  \def\9{\rangle}
  \def\pad{\partial}
  \def\tr{{\rm{tr}}}
  \def\half{\mbox{$1\over2$}}
  \def\cH{{\cal H}}

  \def\bp{{\bf p}}
  \def\bq{{\bf q}}
  \def\bx{{\bf x}}

\begin{document}

\title{Energy density and localization of particles}

\author{Daniel R. Terno}
\affiliation{Department of Physics, Technion---Israel Institute of
Technology, 32000 Haifa, Israel}

\begin{abstract}
We consider the use of the energy density for describing a
localization of relativistic particles. This method is consistent
with the causality requirements. The related positive operator
valued measure is presented. The probability distributions for one
particle states are given explicitly.
\end{abstract}
\maketitle

The concept of particle localization has been raising intriguing
questions almost from the onset of the relativistic quantum
theory. On the one hand, while there is a broad consensus on what
the localization formalism should be and how it can be used, the
formalism itself is still incomplete. Different authors propose
different quantities to describe  localization. On the other hand,
detection events are the experimental basis of particle physics, a
notion of local observables  is a cornerstone of the algebraic
field theory
\cite{ha} of and the coincidence analysis plays an important role
in the general quantum field theory
\cite{ha,qft:b} and quantum optics \cite{mw95}. Moreover, the very concept of
particles takes somewhat a `nebulous character' \cite{bd} in the
field theory in a curved spacetime. A way to substantiate the
notion of particles and to reconcile it with the local framework
of quantum field theory lies in the analysis of the (model)
detectors' excitations \cite{bd,wa:q}. Additional motivation for
deeper understanding of  localization comes from the relativistic
quantum information theory
\cite{pt03}. Usually  information is encoded into discrete degrees of freedom,
but its processing and retrieval are performed in certain places
in space and at certain intervals in time. They are localized, so
there should be a way to describe this localization.

The origins of the problem can be traced to the fact that not only
is there no spacetime localization operator \cite{wi62}, but also
no unique spatial position operator. Moreover, the spatial
position operator  does not exist for massless particles with the
spin larger than $\half$. Even when it does  exist, there may be
no probability current, or causality problems are present
\cite{wi62,nw49,hg74, ali98}.

  In this Letter we investigate the possibility of particle's localization by
means of its energy density, taking as an example a free scalar
boson field. We give an explicit construction of the positive
operator valued measure (POVM) that describes a localization and
derive resulting probability distributions. The limitation of this
approach are discussed in the last part of the Letter. We work in
the Minkowski spacetime, and set $g^{00}=+1$
 and $\hbar=c=1$.

It is  known that particles
  cannot be confined into a volume with a typical dimension  smaller than
  \beq
  \sqrt[3]\Delta> \frac{1}{\6E\9},
  \eeq
   where $\6E\9$ is particle's expected energy \cite{lan}. Therefore, what is actually
  used is
   not an
  operator and its generalized eigenstates, but a  projection valued measure
  that
  is associated with it.

At least some  of the  localization problems can be solved if `observables' and
their spectral measures  are abandoned for the generalized measurements that
are described by POVMs
\cite{ali98,bu99}. A POVM constitutes a non-orthogonal decomposition of the
identity by means of positive operators \cite{povm}.  It is the
most general method to associate probability distributions with
density operators in quantum theory. Nevertheless, some open
questions still remain and it is not easy to derive explicit
expressions from the general principles of covariance, causality,
etc.
\cite{bu99,bu01}.

  Problems with position operators and POVMs  have led to the construction
of various alternative densities and currents. One of the popular
approaches is to use the energy density, i.e., an expectation
value of the renormalized stress-energy tensor
${T}_{00}(\bx,t)={\cH}(\bx,t)$. Here ${T}_{\mu\nu}$ is a
stress-energy tensor, the Hamiltonian  is ${H}=\int{\cH}d^3\bx$
and in the Minkowski spacetime without additional boundary
conditions the renormalization is achieved by  normal ordering.

Consider a one-particle state
  \beq
  |\Psi\9=\int d\mu(p)\psi(\bp)|\bp\9,\qquad
  d\mu(p)=\frac{d^3\bp}{(2\pi)^32p^0},
  \eeq
where $p^0=E(\bp)=\sqrt{m^2+\bp^2}$,
$\6\bp|\bq\9=(2\pi)^3(2p^0)\delta(\bp-\bq)$, and $\int d\mu(p)
|\psi|^2 =1$.
   The inner product of two states is
  calculated as the inner product of their wave functions in  the momentum
  representation,
  \beq
  \6\Psi|\Phi\9=\int d\mu(p)\psi^*(\bp)\phi(\bp).
  \eeq
A wave function in the configuration  space is defined by a
generalized Fourier transform of $\psi(\bp)$ as
  \beq
  \psi(\bx,t)=\int d\mu(p)\psi(\bp)e^{i(\bp\cdot\bx-Et)}.
  \eeq
It is a well-known textbook fact that only momentum space wave
function has a probability interpretation. A construction of
Newton and Wigner \cite{nw49}, which  introduces a new wave
function  \cite{ha}
  \beq
  \psi^{NW}(\bp)=\sqrt{2p^0}\psi(\bp),
  \eeq
is only a partial remedy.  It could not be incorporated into a
continuity equation,
  \cite{ali98, bk01}. If $|\psi^{NW}(\bx,t)|^2$ is taken to be a
  spatial probability distribution  then there are states that violate causality
  \cite{ru87} in the spirit of the Hegerfeldt's theorem that will
  be described below \cite{hg74}.

Energy density is free from these deficiencies. For a state
$|\Psi\9$, it is
\bea
& & E(\bx,t)  =  \6\Psi|{T}_{00}(\bx,t)|\Psi\9   \nonumber \\
&  &=
|\nabla\psi(\bx,t)|^2+|\pad_t\psi(\bx,t)|^2+m^2|\psi(\bx,t)|^2.\label{dens}
\eea
Lorentz transformation properties are built into this quantity by
its definition, and it is positive. Let us consider possible
causality violations. The Hegerfeldt's theorem in its strongest
version proves a superluminal speed for an exponentially localized
particle. If the probability of finding it outside a sphere of
radius $R$ is bounded by
\beq
{\rm Prob}_{\not\in R}< C^2\exp(-2\gamma R),
\eeq
where $C$ is some constant and $\gamma>m$, then the state will
spread faster than light. However, it was shown by Barat and
Kimball \cite{bk01} that no physical state can satisfy this bound.
If $E(\bx,t)$ satisfies it, then both
  $|\psi(\bx,t)|$ and $|\pad_t\psi(\bx,t)|$ are bounded by $C\exp(-\gamma
R)$. It implies that both $\psi(\bp)$ and $\psi(\bp)/E(\bp)$ are analytic
functions in the strip of the complex plane that is bounded by at least $|{\rm
Im}(\bp)|\leq m$, which is inconsistent with the branch cuts in $E(\bp)$ at
$|\bp|=\pm im$. Therefore, the energy density cannot be `localized' enough to
violate
  causality.

  Probability to find a particle in  the spatial region $\Delta$ that is
  centered
  at the point $\bx$ may be considered  a normalized energy density
  \beq
  P_\Delta(\bx,t)=\int_\Delta\frac{E(\bx,t)}{{\6\Psi|{H}|\Psi\9}}d^3\bx.
  \label{prob}
  \eeq
  This quantity still cannot be taken as a probability, because it is
  inconsistent with linearity of the quantum theory. Consider a state
$\rho$ which is a mixture of  two pure states $|\Psi_1\9$ and
$|\Psi_2\9$,
  \beq
  \rho=\alpha \rho_1+(1-\alpha)\rho_2.
  \eeq
According to Eq.~(\ref{prob}) probability densities for the states
$\rho_i=|\Psi_i\9\6\Psi_i|$ are
  \beq
  p_i(\bx,t)=\frac{\6\Psi_i|\cH(\bx,t)|\Psi_i\9}
  {\6\Psi_i|H|\Psi_i\9}=
  \frac{\tr\,\rho_i \cH(\bx,t)}{\tr\,\rho_i H}.
  \eeq
  It is well-known that the probability domain is convex (see, e.g.,
  \cite{qft:b,pt98l}). Therefore,
  \beq
  p_\rho(\bx,t)=\alpha p_1(\bx,t)+(1-\alpha)p_2(\bx,t)\label{conv}.
  \eeq
  On the other hand,
\bea
  p_\rho(\bx,t) & = & \frac{\tr\,\rho\cH(\bx,t)}{\tr\,\rho H}
  \nonumber \\
 & = & \frac{\alpha\tr\,\rho_1 \cH(\bx,t)+(1-\alpha)\tr\,\rho_2 \cH(\bx,t)}
  {\alpha\tr\,\rho_1 H +(1-\alpha)\tr\,\rho_2 H},\label{contra}
  \eea
and these two expressions are generally different.

We can trace the difficulty to the fact that the most general way
to obtain probabilities from density operators is by the means of
POVM, where a normalization of probability follows from the
normalization of a measure. A `manual' normalization, as in
Eq.~(\ref{prob}) brings the nonlinearity that is revealed in
Eq.~(\ref{contra}). However, it is easy to find a remedy. Since
  Hamiltonian densities  commute for  spacelike separated points,
  \beq
  [\cH(x),\cH(x')]=0, \qquad (x-x')^2\leq 0 \label{com}
  \eeq
  the Hamiltonian commutes with the Hamiltonian density at each point.
Therefore, we can construct a POVM element as
  \beq
  {A}(\bx,t)={H}^{-1/2}{\cH}(\bx,t){H}^{-1/2}.\label{povm}
  \eeq
This formal expression glosses over a number of technical points.
First, $H^{-1}$ is ill-defined. However, its action is well
defined when we restrict the Hamiltonian to the non-vacuum states.
We also face a serious problem  in that while ${\cH}(\bx,t)$ is
positive when restricted to the one-particle states, it is not
generally so. We address this issue at the end of this Letter.

Calculation $\6\Psi|{A}(\bx,t)|\Psi\9$ leads to the result which
is similar to the energy density from Eq.~(\ref{dens})
\bea
& & p(\bx,t) =  \6\Psi|{A}(\bx,t)|\Psi\9 \nonumber \\
& & = |\nabla\tilde{\psi}(\bx,t)|^2+|\pad_t\tilde{\psi}(\bx,t)|^2
  +m^2|\tilde{\psi}(\bx,t)|^2 \label{newp},
  \eea
  where an additional $E^{-1/2}$ factor  is added to $\psi(\bp)$,
  \beq
  \tilde{\psi}(\bp)=\frac{\psi(\bp)}{\sqrt{E(\bp)}},
  \eeq
  and
  \beq
  \tilde{\psi}(\bx,t)=\int
  d\mu(p)\tilde{\psi}(\bp)e^{i(\bp\cdot\bx-Et)}.
  \eeq
Working in the momentum space representation, it is easy to see
that indeed
  \beq
  \int p(\bx,t)d^3\bx=1.
  \eeq
   The arguments of Barat and Kimball
   for the energy density are equally well applied to our probability
  distribution $p$, so Hegerfeldt's theorem does not present a paradox.
  Operators
  ${A}(\bx,t)$ have another nice trait: local commutativity, which is a
  desired property for a localization POV measures \cite{bu99}.  Since we
work at the fixed time, two spatially separated regions $\Delta_1$ and
$\Delta_2$,
  $\Delta_1\cap\Delta_2=\emptyset$ are mutually spacelike. We define a POVM
that gives a probability for a particle to be localized at the spatial region
$\Delta_i$ as
\beq
{A}(\Delta_i,t)=\int_{\Delta_i}{H}^{-1/2}{\cH}(\bx,t) {H}^{-1/2}d^3\bx.
\label{domain}
\eeq
  Then Eq.~(\ref{com}) ensures that
  \beq
  [{A}(\Delta_1,t),{A}(\Delta_2,t)]=0
  \eeq
  It is interesting to note that while
  $\tilde{\psi}(\bp)=2^{-1/2}E(\bp)^{-1}\psi^{NW}(\bp)$,  its position
  representation is used in a different manner. Moreover, for the
  states with a well-defined energy, $\Delta E/E\ll 1$, the
  `correct'
  result of Eq.~(\ref{povm}) and the `wrong' one of Eq.~(\ref{prob}), augmented
  with the prescription of Eq.~(\ref{conv}) to handle mixed states give
  nearly identical probability distributions.

The POVM of Eq.~(\ref{domain}) satisfies the standard list of
requirements for localization operators \cite{bu99} and solves a
problem of localization in one-particle theory. However, the
notion of localization that is based on the energy density cannot
have a universal validity. There is no probability density `in
general', but only a probability density that is related to a
specific detection scheme.

A simple example when it runs afool is the Unruh effect \cite{bd,
wa:q,ueff}. An accelerated detector that moves in the Minkowski
vacuum responds as an inertial detector would if immersed into a
thermal bath of the temperature
\beq
  T=a/(2\pi k_B),
\eeq
where $k_B$ is Boltzmann's constant and $a$ the proper
acceleration. A perturbative derivation of this result is
supported by the Green's function analysis. Green functions for
the accelerated detector are identical with the thermal Green
functions of the inertial one at the  temperature $T$. However,
the expectation of the renormalized stress-energy tensor is zero
in {\em both} inertial and accelerated frames.

  In a more complicated settings a question of positivity becomes acute.
Classical energy density is always positive, which is to say that
the stress-energy tensor for scalar field satisfies the weak
energy condition (WEC)
  $T_{\mu\nu}u^\mu u^\nu\geq 0$, where $u^\mu$ is a causal vector.
In the framework of  general quantum field theory \cite{ha,qft:b} it is
impossible. There are states $|\Upsilon\9$ that violate WEC, namely
$\6\Upsilon|{T}_{\mu\nu}u^\mu u^\nu|\Upsilon\9 \leq 0$ holds
\cite{nc65}, where $T_{\mu\nu}$ now is a renormalized stress energy operator.
 For example, squeezed states of electromagnetic
\cite{mw95} or scalar field have negative energy densities \cite{sq,f01}.
  It is known that even if  WEC is violated  the average WEC
  \beq
  \int_{-\infty}^{\infty}{T}_{\mu\nu}u^\mu u^\nu d\tau\geq 0,
  \eeq
still holds when the integral is taken over the world line of a
geodesic observer (inertial observer in the Minkowski spacetime)
\cite{tip}. There are also more stringent quantum inequalities
that limit the amount of the WEC violation. Instead of infinite
time interval they deal with a sampling that is described by a
function with a typical width $t_0$ \cite{f01}.  Behaviour of
fields  subjected to boundary conditions is more complicated, but
similar constraints exist also in these cases
\cite{f01}.
   To our
ends we need the analogous inequalities to hold for a spatial averaging. This
is, however, impossible. A class of quantum states was constructed for a
massless, minimally coupled free scalar field  (superposition of the vacuum and
multi-mode two-particle states). These states can produce an arbitrarily large
amount of negative energy in a given finite region of space at a fixed time
 \cite{f02}. In this and similar cases the  spatial averaging over
  part of a constant time surface does not produce a positive quantity.
  Consequently, the probabilistic interpretation of the energy density
  fails.

\medskip

It is a pleasure to thank Netanel Lindner and Asher Peres for
useful discussions and their critical comments. This work is
supported by a grant from the Technion Graduate School.
  
  \end{document}